\documentclass[12pt,preprint]{aastex}
\def\comma{,~}
\begin{document}

\title{The high-energy gamma-ray fluence and energy spectrum of GRB 970417a from observations 
with Milagrito}

\author{
R.~Atkins,\altaffilmark{1}
W.~Benbow,\altaffilmark{2}
D.~Berley,\altaffilmark{3,10}
M.L.~Chen,\altaffilmark{3,11}
D.G.~Coyne,\altaffilmark{2}
B.L.~Dingus,\altaffilmark{1}
D.E.~Dorfan,\altaffilmark{2}
R.W.~Ellsworth,\altaffilmark{5}
D.~Evans,\altaffilmark{3}
A.~Falcone,\altaffilmark{6,12}
L.~Fleysher,\altaffilmark{7}
R.~Fleysher,\altaffilmark{7}
G.~Gisler,\altaffilmark{8}
M.M.~Gonzalez Sanchez,\altaffilmark{1}
J.A.~Goodman,\altaffilmark{3}
T.J.~Haines,\altaffilmark{8}
C.M.~Hoffman,\altaffilmark{8}
S.~Hugenberger,\altaffilmark{4}
L.A.~Kelley,\altaffilmark{2}
S.~Klein,\altaffilmark{2,13}
I.~Leonor,\altaffilmark{4,14}
J.F.~McCullough,\altaffilmark{2,15}
J.E.~McEnery,\altaffilmark{1}
R.S.~Miller,\altaffilmark{8,6}
A.I.~Mincer,\altaffilmark{7}
M.F.~Morales,\altaffilmark{2}
P.~Nemethy,\altaffilmark{7}
J.M.~Ryan,\altaffilmark{6}
F.W.~Samuelson,\altaffilmark{8}
B.~Shen,\altaffilmark{9}
A.~Shoup,\altaffilmark{4}
C.~Sinnis,\altaffilmark{8}
A.J.~Smith,\altaffilmark{9,3}
G.W.~Sullivan,\altaffilmark{3}
T.~Tumer,\altaffilmark{9}
K.~Wang,\altaffilmark{9,16}
M.O.~Wascko,\altaffilmark{9,17}
S.~Westerhoff,\altaffilmark{2,18}
D.A.~Williams,\altaffilmark{2}
T.~Yang,\altaffilmark{2}
G.B.~Yodh\altaffilmark{4} \\
}

\altaffiltext{1}{University of Wisconsin, Madison, WI\,53706, USA}
\altaffiltext{2}{University of California, Santa Cruz, CA\,95064, USA}
\altaffiltext{3}{University of Maryland, College Park, MD\,20742, USA}
\altaffiltext{4}{University of California, Irvine, CA\,92697, USA}
\altaffiltext{5}{George Mason University, Fairfax, VA\,22030, USA}
\altaffiltext{6}{University of New Hampshire, Durham, NH\,03824, USA}
\altaffiltext{7}{New York University, New York, NY\,10003, USA}
\altaffiltext{8}{Los Alamos National Laboratory, 
                 Los Alamos, NM\,87545, USA}
\altaffiltext{9}{University of California, Riverside, CA\,92521, USA}
\altaffiltext{10}{Permanent Address: National Science Foundation, Arlington, VA\
,22230, USA}
\altaffiltext{11}{Now at Oak Ridge National Laboratory, Oak Ridge, TN\, 37831-6203, USA}
\altaffiltext{12}{Now at Purdue University, West Lafayette, IN\, 47907, USA}
\altaffiltext{13}{Now at Lawrence Berkeley National Laboratory, Berkeley, CA 94720, USA}

\altaffiltext{14}{Now at University of Oregon, Eugene, OR\, 97403, USA}
\altaffiltext{15}{Now at Cabrillo College, Aptos, CA\, 95003, USA}
\altaffiltext{16}{Now at Armillaire Technologies, Bethesda, MD\, 20817, USA}
\altaffiltext{17}{Now at Louisiana State University, Baton Rouge, LA\, 70803, USA}
\altaffiltext{18}{Now at Columbia University, New York, NY\, 10027, USA}

\begin{abstract}
Evidence of TeV emission from GRB970417a has been previously reported
using data from the Milagrito detector~\cite{atkins00b}. Constraints on 
the TeV fluence and the energy spectrum are now derived using additional 
data from a scaler system that recorded the rate of signals from the Milagrito
photomultipliers. This analysis shows that if emission from GRB970417a
has been observed, it must contain photons with energies above 650
GeV.  Some consequences of this observation are discussed.

\end{abstract}

Subject headings: gamma rays -- bursts, observations

\section{Introduction} 

Some of the most important contributions to our understanding of
gamma-ray bursts (GRBs) have come from observations of afterglows over
a wide spectral range. Comparisons between these observations and
predictions of GRB afterglow properties both as a function of time and
of wavelength have provided stringent tests of GRB
models~\cite{paradijs00}. However, far less is known about the
multiwavelength spectrum during the prompt phase of GRBs because of
its very short duration.

Almost all GRBs have been detected in the energy range between 20 keV
and 1 MeV~\cite{fishman95}. A few have been observed above 100 MeV by
EGRET~\cite{schneid92,sommer94,hurley94,schneid95} indicating that at
least some GRB spectra extend up to hundreds of MeV. However, the
upper extent of GRB energy spectra is unknown. There may be a second
(higher energy) component of emission, similar to that seen in several
TeV sources~\cite{dermer99,pilla98}.

The Milagro gamma-ray observatory, which began full operations in
January 2000, is a wide field-of-view instrument that operates with a
duty cycle near 100 \%. It is particularly well suited to extending
observations of the prompt phase of GRBs up to TeV energies. A
prototype of Milagro, Milagrito~\cite{atkins00a}, found evidence for
TeV emission from one of the 54 gamma-ray bursts observed by BATSE
that were within the Milagrito field of view with a probability of
$1.5 \times 10^{-3}$ of being a background
fluctuation~\cite{atkins00b}.  An excess of events was observed from
the direction of GRB 970417a during the time BATSE observed
emission. In this paper, we use additional data from Milagrito to
examine the fluence and spectrum of the possible high-energy emission
of GRB970417a.

Milagrito detected secondary particles reaching the ground produced by
the interaction of TeV gamma rays in the atmosphere. The Milagrito
detector consisted of a single layer of 228 photomultiplier tubes
(PMT) placed on a 2.8 m $\times$ 2.8 m grid approximately 1 meter
below the surface of a large covered pond of
water~\cite{atkins00a}. An air shower was registered when relativistic
charged particles, radiating Cherenkov light in the water, caused $\ge$
100 PMTs to detect light within a 200-ns time interval.  The direction
of the gamma ray initiating the shower was reconstructed from the
relative timing of the PMT signals.

It is very difficult to obtain information on the energy of the
individual events contributing to the TeV gamma-ray excess observed
with Milagrito. While the observed number of shower particles is
related to the energy of the primary gamma-ray, it also depends on the
height in the atmosphere of the first interaction and on the distance
of the shower core from the pond. The area of Milagrito was small
relative to the lateral extent of a typical shower, so it is usually
not possible to determine the location of the core.  Consequently, the
number of PMTs hit in the pond is only weakly related to the energy of
the gamma-ray primary. Also, the trigger required 100 of the 228 PMTs
to register a signal, so there is very little dynamic range over which
to identify a variation due to different source spectra.

Instead, information about the distribution of the gamma-ray energies
can be obtained using additional information about the summed count
rates of the individual PMTs in the pond. Very-high-energy (VHE)
gamma rays of too low an energy to trigger the detector can give rise
to an increase in these rates. Measurement of the rates therefore
allows constraints to be placed on the spectrum of the putative TeV
gamma-ray flux observed from GRB970417a.

\section{Observations with the Triggered Milagrito Data}

The angular resolution of Milagrito was substantially better than that
of BATSE. Thus it was decided {\it a priori} to perform a search for a
TeV excess everywhere within the area of the BATSE 90\% positional
uncertainty (statistical+systematic)~\cite{briggs99} for the T90
period reported by BATSE ($7.9$~s for GRB970417a). The search area
from BATSE was tiled with an overlapping array of 1.6$^{\circ}$ radius
circular bins (appropriate for the angular resolution of
Milagrito) centered on points on a $0.2^{\circ} \times 0.2^{\circ}$ grid. The
Poisson probability that the observed number of events was a
fluctuation of the background was calculated for each bin.  The bin
with lowest probability of being a background fluctuation for GRB970417a 
contains 18 events.  In the null hypothesis (in which we assume
there is no signal), the best estimate of the background is $3.46 \pm
0.11$ events.  The Poisson probability of observing 18 events in a
given bin with the background level of 3.46 is $2.9 \times
10^{-8}$. After accounting for the oversampling in this method, the chance 
probability of such an observation anywhere
within the search region for GRB970417a is $2.8 \times
10^{-5}$~\cite{atkins00b}. Since 54 gamma-ray bursts were examined for 
VHE emission, the probability that this excess was a chance coincidence 
was 1.5$\times 10^{-3}$.

If we assume that the observation of GRB970417a is due to signal plus
background events, then the best estimate of the background level in
the signal region is different from that obtained assuming the null
hypothesis. This is because the search method will tend to select the
bin with an upward fluctuation of the background in addition to the
signal. A Monte Carlo simulation was performed to find the average
number of signal events that results in an excess of 18 events in the
most significant bin, given a mean background level of 3.46. The
result is that the average number of signal events in the most
significant bin is $13.3\pm 4.2$ in this case: the remaining 4.7
events are background.

\section{Observations using the Milagrito Scaler System}
The Milagrito scaler system recorded the PMT counting rates once per
second for two brightness thresholds. Low threshold, which is the rate at
which the PMTs detected $\sim$0.25 photoelectrons or more, and high
threshold, which is the rate at which the PMTs detected $\sim$ 5 or
more photoelectrons.  Here we discuss the results from the scalers
recording the PMT rates above the low-discriminator threshold (about
0.25 photoelectrons).  The PMT signals were processed in groups of 16
by the electronics~\cite{atkins00a}.  Each of the 16 PMTs in a group
was connected to one of four scalers. The connections were arranged so
that no pair of nearest or diagonal neighbor PMTs were connected to
the same scaler. Each of these scaler outputs was a logical OR of the
signals from four PMTs. Consequently, if two or more PMTs connected to
the same scaler were hit within about 30 ns of each other, only one
output count was generated.

\begin{figure}
\plotone{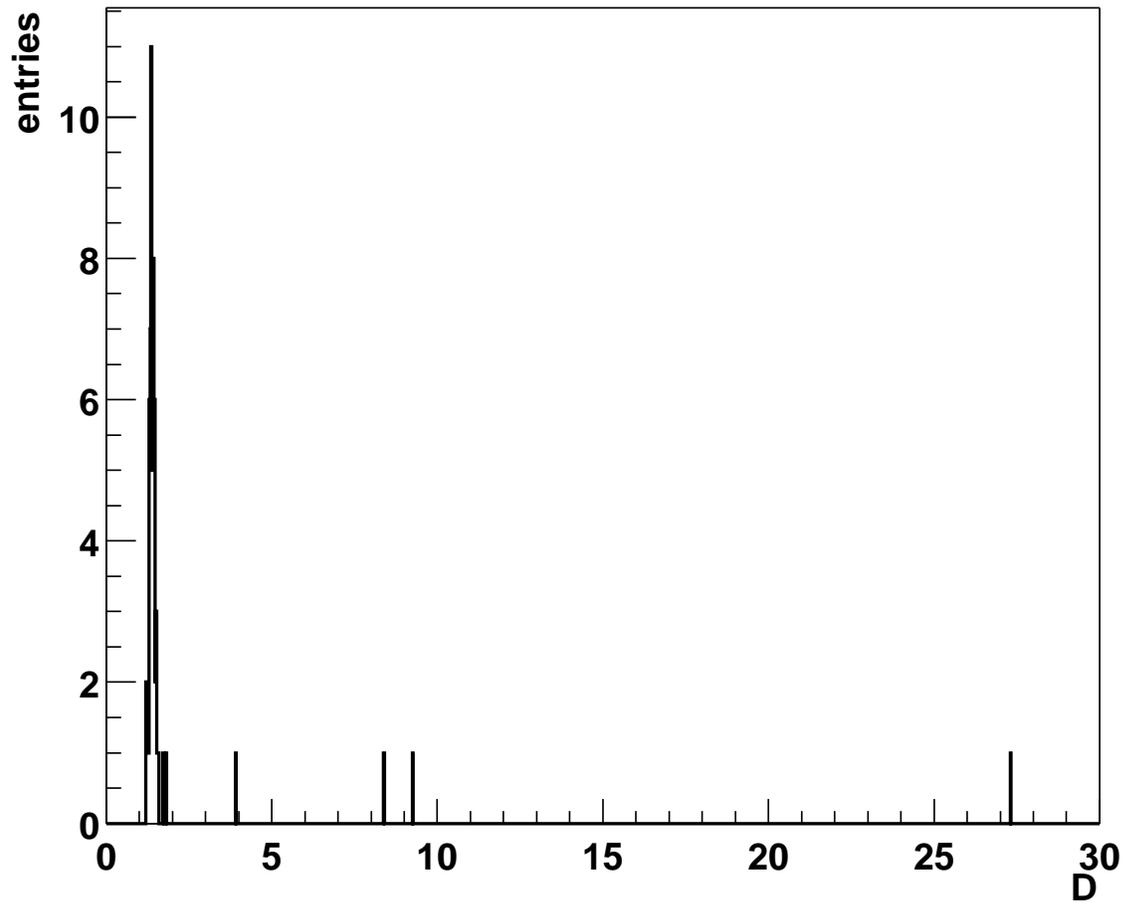}
\caption{The distribution of D, the noise in the scaler rates, for the
60 low threshold scaler channels; the four channels clearly more noisy
that the rest were discarded.}
\label{fig:chstats}
\end{figure}

The analysis of the scaler data involves a search for a significant
increase in the PMT scaler rates at the time of GRB970417a. Noisy
channels were eliminated from the analysis to improve the sensitivity
of the PMT singles-rate analysis. There were a few PMTs that were
noisy and several PMTs that had large rate variations due to pinhole
light leaks in the pond cover.  GRB970417a occurred 82 minutes after
sunrise at the Milagro site, so that light leaks cannot be ignored.
Noisy PMTs were identified by examining the rms fluctuations in the
count rates of each scaler channel. Figure~\ref{fig:chstats} shows the
distribution of $D= \frac{rms}{\sqrt{<rate>}}$ for the 60 scaler
channels, for an hour-long period centered on the burst trigger time.
$D\sim1$ for channels in which the scaler counts follow a Poisson
distribution, while noisy channels will have $D >>$ 1. The four
channels with $D >$ 2 were excluded from further analysis.  The
remaining channels have $<D>=1.4$, somewhat larger than is expected
for Poisson statistics. 

\begin{figure}[tbh!]
\plotone{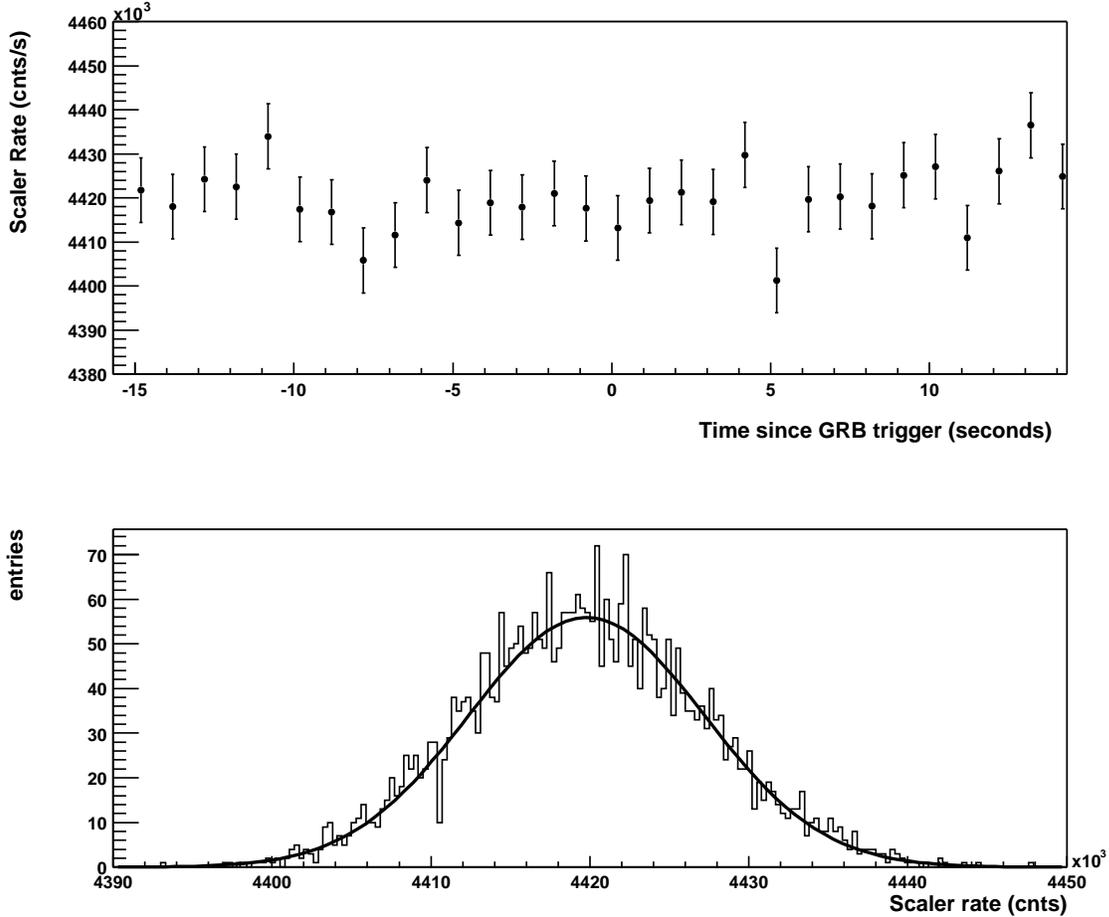}
\caption{Top panel: the summed scaler rate in one second bins around
the burst time. Bottom panel: the distribution of summed scaler rates
for one hour around the burst trigger time.  Superimposed on the data
is a fit to a Gaussian.}

\label{fig:dist1s}
\end{figure}

The 56 scaler channels without excessive noise were added
together to give the summed scaler rate as a function of
time. Figure~\ref{fig:dist1s} shows the summed scaler rate for the
30-second interval centered on the start time of GRB970417a and the
distribution of values of the summed scaler rate for 60 minutes around
the time of the burst. No increase in the summed scaler rate is
apparent at the time of GRB970417a.

\begin{figure}[tbh!]
\plotone{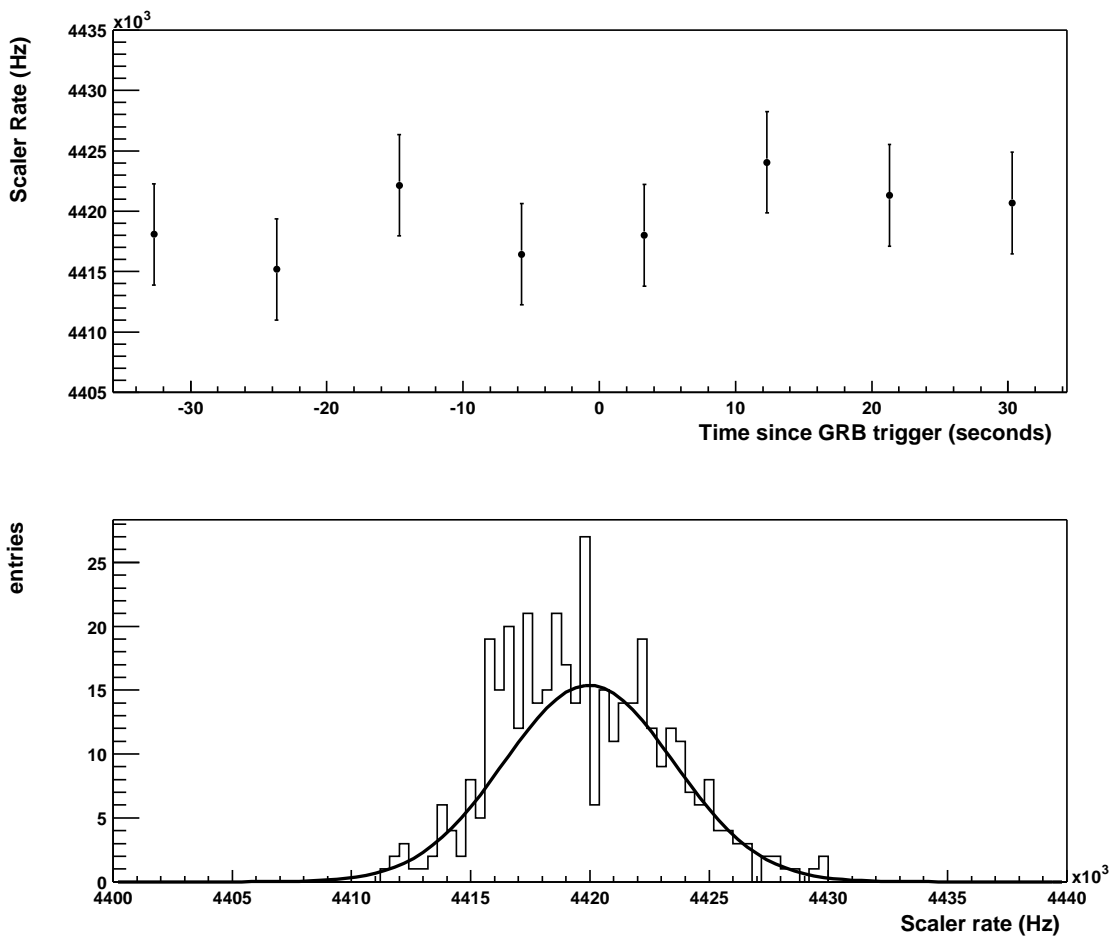}
\caption{Top panel: the summed scaler rate in 9 second bins around the
burst time. Bottom panel: the distribution of summed scaler rates for
one hour around the burst trigger time. The solid line shows a fit to
a Gaussian, which has a mean of 4.420 MHz and a rms width of 4100 Hz.}
\label{fig:dist8s}
\end{figure}

We wish to look for an excess in the summed scaler rate during T90, as
this was the interval over which the triggered excess was observed in
Milagrito. The BATSE trigger time for GRB970417a was 50016.71 s on MJD
50555: the start of the T90 interval was offset by -1.024 s from the
trigger time, and the duration of T90 was $7.9$~s. It is necessary to
examine a 9-second time bin to fully contain the T90 interval as the
scaler data were recorded once per second.  Figure~\ref{fig:dist8s}
shows the summed scaler data for 9-second intervals.

The summed scaler rate during the 9 seconds that contained the burst
was 4.418 MHz, which is 0.5 standard deviations below the mean of
4.420 MHz. Thus there is no evidence for an increase in the summed
scaler rate during T90 of GRB970417a.  The 99.9\% confidence level
upper limit on the number of excess summed scaler counts due to gamma
rays during T90 of GRB970417a has been calculated using the method
described in Helene (1988), assuming that the distribution of summed
scaler rates due to background is a Gaussian with the mean and
standard deviation shown in Figure~\ref{fig:dist8s}. This background
distribution is significantly broader than expected from Poisson
statistics due to correlations between hits. Monte Carlo studies show
that this distribution is expected to be broadened due to correlations
between the scaler channels as several PMTs can detect light from a
single shower particle.  This distribution also incorporates
systematic fluctuations at a 9 second time scale.  Using the above
distribution to calculate the upper limit, these effects are taken
into account.  The result is that fewer than 67000 excess counts from
the scalers were detected over T90 of the burst. This quantity is
defined as $N_{upper}$(scaler).

\section{Sensitivity of the Milagrito Detector}

The scalers could register counts from showers that were too low in
energy to trigger 100 PMTs in Milagrito.  Thus the observation of an excess in the
triggered data but not in the summed scaler rate would imply that the
observed excess is not due solely to a large flux of low-energy gamma
rays. A simulation of the energy-dependent response of the Milagrito
trigger and of the summed scaler rate is needed to quantify the
constraints on the gamma-ray flux.  The simulation reflected the
instrumental conditions at the time of the burst, including the fact
that 9 PMTs were not operational at that time.  The four noisy scaler
channels were also discarded in the simulation of the scalers.  For
simulations of the triggered Milagrito data (summed scaler rate),
showers were thrown at discrete energies from 50 (10) GeV to 50 TeV 
at zenith angles from 20-25 degrees. GRB970417a was at a
zenith angle of 22 degrees for Milagrito.

\begin{figure}[tbh!]
\plotone{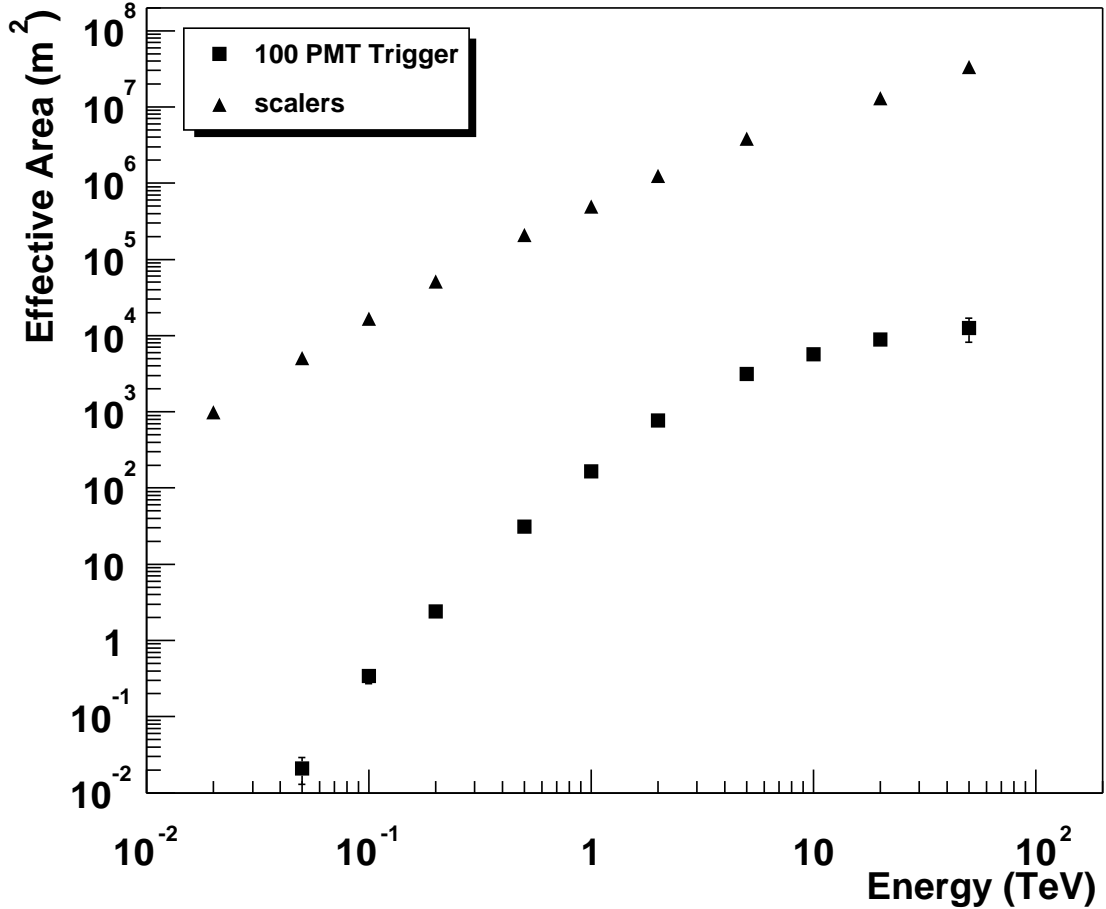}
\caption{The effective area for the Milagrito trigger (squares) and 
summed scaler rate (triangles) as a function of incident energy for gamma rays at a zenith 
angle of 22 degrees}.
\label{fig:area}
\end{figure}

If {$n_{total}$} simulated showers are thrown with cores uniformly
distributed over a sufficiently large area, $A_{core}$, so that there
is only a negligible probability of showers outside that area
triggering the detector, then the effective area, $A_{trigger}$, is given
by:

\begin{equation}
A_{trigger} = { A_{core} \times \frac {n_{trigger}}  {n_{total}}},
\end{equation}
where {$n_{trigger}$} is the number of simulated events that would
trigger the detector.  Similarly an effective area for the summed
scalers is defined by:

\begin{equation}
A_{scaler} = { A_{core} \times \frac {n_{scaler}} {n_{total}}},
\end{equation}
where {$n_{scaler}$} is the number of simulated scaler counts that
would be recorded by the detector.  Note that more than one scaler
count can result from a single shower.  Table~\ref{tab:area} and
Figure~\ref{fig:area} give the effective areas for triggers and
scalers as functions of incident gamma-ray energy.  Each simulated 
shower was used only once in deriving the
effective areas.

\begin{table}[!h]
\begin{tabular}{lll}
Energy (GeV) &$A_{trigger}$ (m$^2$) &$A_{scaler}$ (m$^2$) \\
 10 & - & $2.88 \pm 0.12 \times 10^2$ \\
 20 & - & $9.79  \pm 0.61 \times 10^2 $ \\
50          & $0.021 \pm 0.08$  &  $5.10 \pm 0.08 \times 10^3$ \\
100         & $0.34  \pm 0.07$ &  $ 1.67 \pm 0.09 \times 10^4$ \\
200         & $2.4 \pm 0.35$ &   $5.12  \pm 0.20 \times 10^4$ \\
500         & $31.4 \pm 3.0$ &  $2.09  \pm 0.09 \times 10^5$ \\
1000        & $166.3 \pm 6.1$ &  $4.90 \pm 0.19 \times 10^5$  \\
2000        &  $768 \pm 48$   &   $1.24  \pm 0.06 \times 10^6$\\
5000        &  $3139 \pm 124$  &   $3.80  \pm 0.22 \times 10^6$\\
10000       &  $5726 \pm 254$   &   $7.03 \pm 0.50 \times 10^6$  \\
20000       &  $8888  \pm 1000$ &  $1.21  \pm 0.17 \times 10^7 $\\
50000       &  $12549 \pm 4447$  &  $3.32 \pm 0.36 \times10^7$     \\
\end{tabular}
\caption{Milagrito triggered and scaler effective area vs. energy for gamma rays at a zenith
angle of 22 degrees from Monte Carlo studies. The number of triggers that
reconstruct within 1.6 degrees of the true incident direction, is used
to calculate the triggered effective area. }
\label{tab:area}
\end{table}

\begin{figure}[tbh!]
\plotone{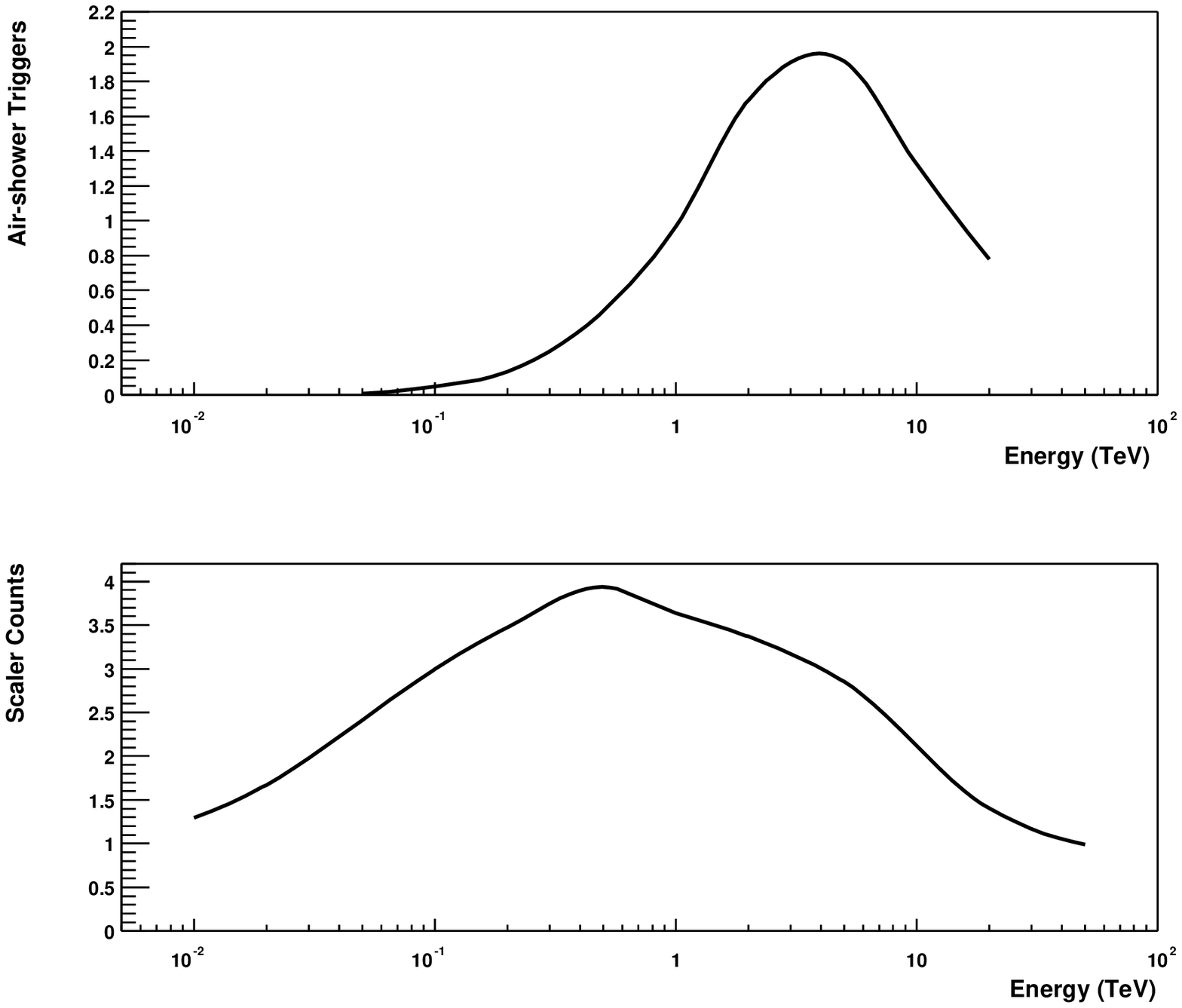}
\caption{Milagrito energy response to an E$^{-2.4}$ differential photon spectrum.
The top panel is for triggered events and the lower panel is for the 
summed scalers. The vertical scales for each panel are arbitrary. This figure
illustrates that if one were to introduce and reduce an upper energy cutoff, the
ratio of excess scaler counts to air-shower triggers increases.}
\label{fig:trig_en}
\end{figure}

If the effective area is convolved with a test input spectrum, the
result is the expected distribution of gamma-ray energies for events
detected by Milagrito for that spectrum.  Figure~\ref{fig:trig_en}
shows this distribution for triggers (top) and the summed scalers
(bottom) assuming an input gamma-ray spectrum with a differential
power law index of -2.4 and no upper energy cutoff. The scaler
response extends to considerably lower energies, so that the
contribution of the scalers becomes enhanced relative to the triggered
data as one introduces and reduces an upper energy cutoff.

\section{Constraints on photon energies and the fluence from GRB970417a}

The true spectrum of the gamma rays responsible for the observed
excess from GRB970417a at TeV energies is unknown. However,
constraints can be derived by requiring the photon energy spectrum and
fluence to be consistent with the Milagrito observations.  In
particular, the spectrum convolved with the Milagrito response
functions should reproduce the observed excess in the triggered data
while not exceeding the upper limit, $N_{upper}(scaler) $, obtained in
the scaler analysis.

\subsection{Monoenergetic Spectrum}
It is instructive to first consider the case of a monoenergetic gamma-ray
spectrum.  The number of triggered events produced by a monoenergetic 
flux, $F(E)$, in a burst lasting T seconds is given by:

\begin{equation}
N_{trig} = F(E) \times T \times A_{trig}(E),
\end{equation}
where $A_{trig}(E)$ is the trigger effective area for incident
gamma rays with energy $E$.
 
Similarly, the number of summed scaler counts produced by the flux
$F(E)$ is given by:

\begin{equation}
N_{scalers} = F(E) \times T \times A_{scalers}(E).
\end{equation}

It is common practice to describe the output of a GRB by its
fluence, which is the integral of the flux over energy and time: thus
the fluence for a steady monoenergetic flux at an energy $E$ is simply
$E \times T \times F(E)$.

\begin{figure}[tbh!]
\plotone{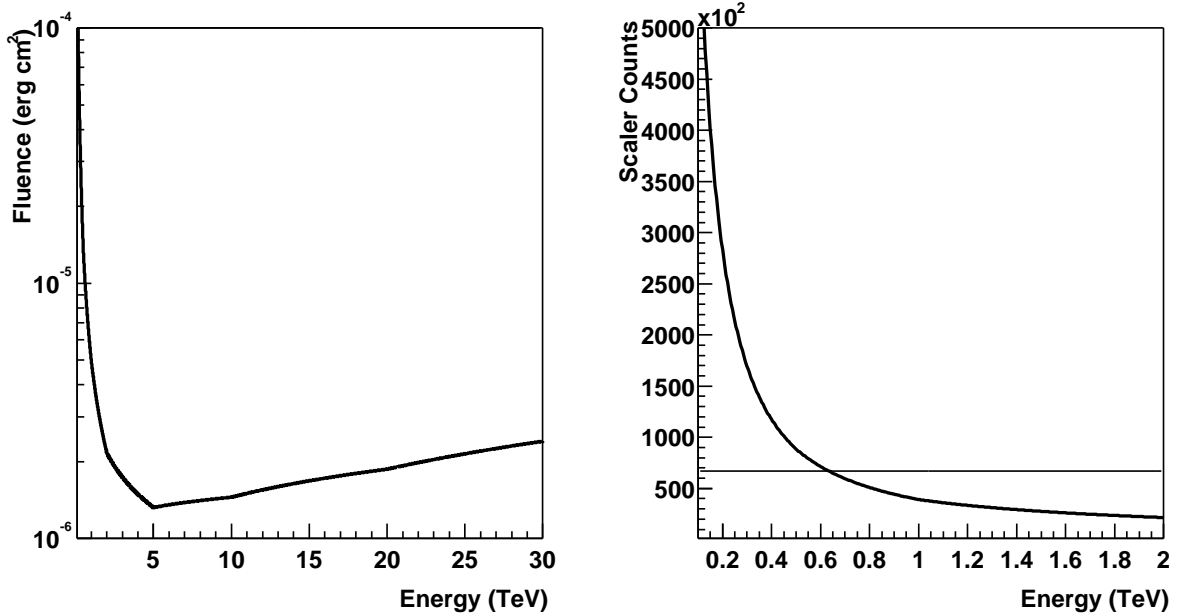}
\caption{The left hand panel shows the fluence implied by the
triggered-event excess observed in Milagrito assuming a monoenergetic
spectrum for different values of the energy. The right hand panel
shows the predicted number of summed scaler counts for a monoenergetic
spectrum for different values of the energy. The horizontal line
indicates the upper limit $N_{upper}$(scaler) $<$ 67000.}
\label{fig:mono}
\end{figure}

Figure~\ref{fig:mono} shows the fluence implied by the triggered-event
excess observed in Milagrito assuming a monoenergetic spectrum for
different energy gamma-rays.  The fluence has a minimum of $1.3 \times
10^{-6}$ ergs/cm$^2$ for a monoenergetic flux at 5 TeV. This is the
minimum fluence that could be responsible for the triggered-event
excess observed in Milagrito regardless of spectral shape.  Also shown
in Figure~\ref{fig:mono} is the expected number of excess summed
scaler counts as a function of energy for a monoenergetic spectrum
with the fluence necessary to produce the observed triggered-event
excess. The horizontal line show $N_{upper}$(scaler)$<$67000, the
99.9\% upper limit on the summed scaler excess derived above.  Because
the summed scaler excess must be less than $N_{upper}$(scaler),
GRB970417a cannot have a monoenergetic flux with any energy below 650
GeV. The Milagrito observation cannot be due solely to gamma rays with
energies below 650 GeV regardless of spectral shape.

\subsection{Truncated power law Spectrum}

In reality GRBs probably have an upper energy cutoff due to the
intrinsic radiation emission mechanism at the
source~\cite{baring97,lithwick01}, attenuation via pair production with
intergalactic photon fields~\cite{primack99,salamon98}, or both.  As an
example of how the analysis of the Milagrito result can be applied to
a more realistic spectral shape, we assume that the photon spectrum is
a truncated power law:

\begin{equation}
\frac {dN} {dE} = cE^{-\alpha}, E_{min} \le  E \le E_{max}
\end{equation}
\begin{equation}
\frac {dN} {dE} = 0, E >  E_{max},
\end{equation}
where N is the number of photons at energy E, $\alpha$ is the spectral index, 
and c is a normalization factor.  
The number of events detected by Milagrito for this spectrum is given by:

\begin{equation}
N_{signal} = \int_{E_{min}}^{E_{max}}A_{eff} \frac {dN} {dE} dE.
\end{equation}

Inserting the Milagrito effective area for the summed scalers
(triggered data) in this equation results in the predicted number of
excess summed scaler counts (triggered events). Since 13.3 excess
triggered events were observed, any combination of $\alpha$, $E_{min}$
and $E_{max}$ can be used in equation (7) to find c (the normalization
constant) as a function of $\alpha$ and $E_{max}$), and thus the
number of predicted summed scaler counts.

We take $E_{min} = 50 GeV$ so that the contribution to the observed
Milagrito triggered event excess from the portion of the spectrum
below this energy is negligible.  We then allow $\alpha$ and $E_{max}$
to vary and search for combinations of $E_{max}$ and $\alpha$ that
satisfy the requirement that the predicted number of summed scaler
counts be less than $N_{upper}$(scaler).  (Strictly speaking there is
contribution to the summed scaler counts from events below 50 GeV, but
this is small for the cases discussed below).

\begin{figure}[tbh!]
\plotone{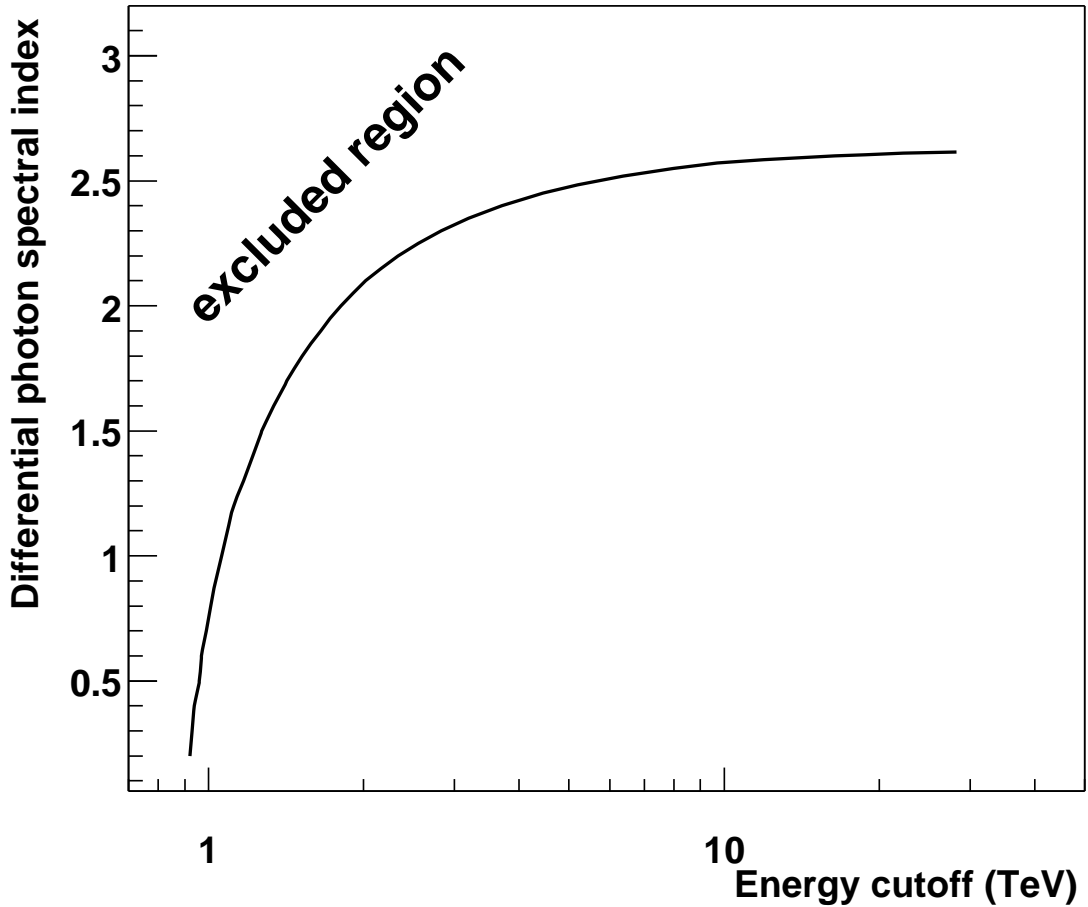}
\caption{The curve shows the combinations of spectral index and upper
energy cutoff that give rise to 67000 excess scaler counts and 13.3
excess triggered events. Regions to the top left of this curve, i.e.
softer spectra and lower energy cutoff, predict $\ge 67000$ excess
scaler counts and are thus excluded by the Milagrito observations.}
\label{fig:excl}
\end{figure}

Figure~\ref{fig:excl} shows the combinations of spectral index and
maximum energy that satisfy the condition that the number of summed
scaler counts equals $N_{upper}(scaler)$.  Regions above and to the
left of this curve (lower energy cutoff or softer spectra) result
in a summed scaler excess greater than 67000, and are thus
inconsistent with the Milagrito observations.  This plot shows that
for a differential photon spectrum $\sim E^{-0.5}$ (a very hard
spectrum), the spectrum must extend up to at least 1 TeV to be
consistent with the Milagrito observations. For a spectral index of
-2.4, more typical of what is observed from other VHE sources, the
spectrum must extend to at least 5 TeV.  Note that photon energies
from GRB970417a must extend to higher energies for a truncated
power-law spectrum than for a monoenergetic spectrum.

The results obtained here incorporate the effects of statistical and
systematic fluctuations in the scaler rate.  However, there may also
systematic uncertainties in the calculation of the detector
response. Previous studies have shown that the Monte Carlo simulation
of the Milagrito detector correctly predicts the trigger rate from
cosmic-ray protons, helium and CNO~\cite{atkins99}. This implies that
the Milagrito effective area for cosmic rays is correctly modeled to
within $\sim15\%$, and that the energy scale of the response of
Milagrito agrees with the simulation to within $\sim10\%$. There may
also be systematic errors associated with simulating shower
proipagation, which could cause an additional small uncertainty for
the flux from GRB970417a due to the triggering efficiency of
cosmic-rays relative to gamma-rays. The simulation of the scaler
response is subject to many of the same systematic errors as the
triggered response of Milagrito.  However, there are additional
systematic uncertainties in the modeling of the scaler data since they
are more sensitive to fluctuations in the tails of the air-shower
distributions. It is difficult to quantify these additional systematic
uncertainties, but we note that seasonal fluctuations of around 15\%
are observed in the scaler rates (but not the trigger rate). These
seasonal fluctuations in the scaler rates are probably due to changes
in the vertical density profile of the atmosphere. Overall we estimate
the systematic uncertainty in the constraints on maximum energy to be
around 35\%.

\section{Comparison with other observations of GRB970417a}

GRB970417a was a weak, soft GRB as observed by BATSE.  The fluence
in the 50--300 keV range was $1.5\times 10^{-7}$ ergs/cm$^2$.   
The low fluence implies that spectral fits to these data are poorly
constrained.  A fit to a broken power law~\cite{band93} yields a
relatively low break-energy of $58 \pm 18$ keV, with an undefined low
energy spectral index and a soft power law above the break with a
spectral index of $-2.67 \pm 0.4$~\cite{connaughton99}.  A power law
with no break fits the BATSE data equally well with a spectral index
of $-2.36\pm 0.2$~\cite{kanekoy00}.  In what follows, we use this
spectral form for this burst, since the need for fewer parameters
results in a more tightly constrained fit.

\begin{figure}[tbh!]
\plotone{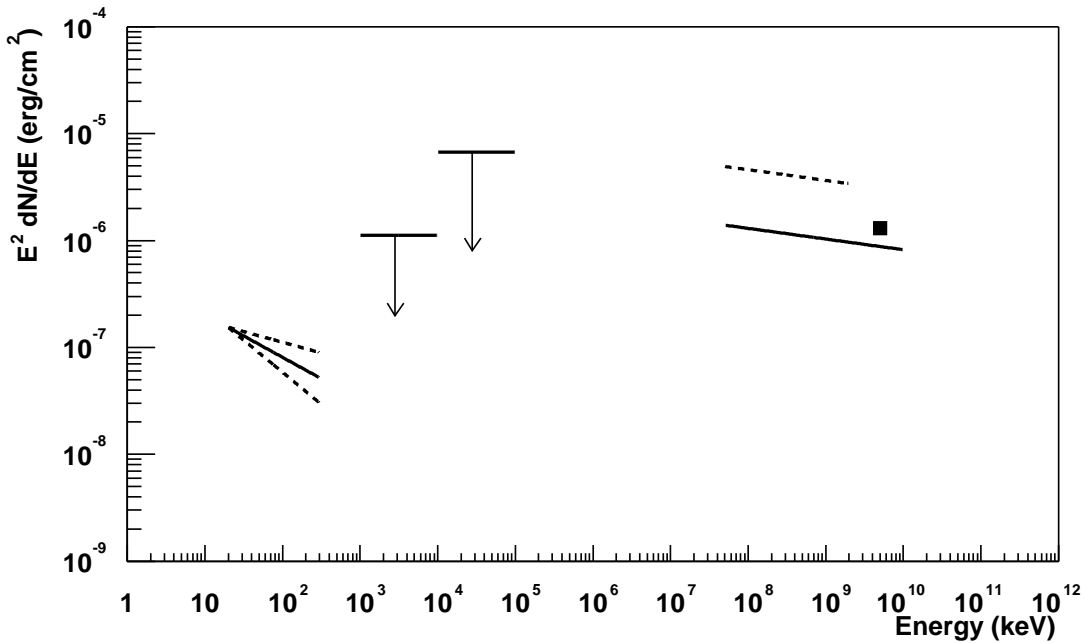}
\caption{The spectral energy distribution for GRB970417a showing a
single power-law fit to the BATSE data, upper limits at 1 MeV and 10
MeV from the EGRET TASC detector and 3 possible spectral forms
consistent with the Milagrito observations.}
\label{fig:nufnu}
\end{figure}

We have calculated upper limits for the fluence in the ranges 1--10
MeV and 10--100 MeV using data from the EGRET Total Absorption
Scintillation Counter (TASC) detector.  Figure~\ref{fig:nufnu} shows
the spectral energy distribution for GRB970417a, including the fit
BATSE spectrum, the upper limits from the EGRET TASC detector, and the
VHE flux implied by the Milagrito observations assuming 3 possible
spectral forms:

1) The square symbol assumes a monoenergetic flux at 5 TeV: this is the lowest VHE
fluence compatible with the Milagrito observations. 

2) The lower (solid) line assumes a power law spectrum with differential photon index of -2.1. 

3) The upper (dashed) line assumes a power-law spectrum with differential photon
index of -2.1 and a sharp cutoff at 2 TeV.

\section{Discussion}

As shown above, if the excess triggered events observed by Milagrito
were indeed associated with GRB970417a, then they must be due to
photons of at least 650 GeV, the highest energy photons detected from
a GRB. Figure~\ref{fig:nufnu} shows that the VHE fluence must be at
least an order of magnitude greater than the sub-MeV fluence measured
by BATSE. In the following, assuming that high-energy emission from
GRB970417a was detected by Milagrito, this section explores some of the
consequences of the observation.

The Milagrito observations at VHE energies are not consistent with a
straightforward extrapolation of the spectrum measured by BATSE, so
they would be the first evidence for the existence of a second (higher
energy) emission component in a gamma-ray burst. We note that while
the implied fluence at VHE energies is considerably larger than that
measured by BATSE, the actual peak of the spectral energy distribution
may lie below the energy range measured by BATSE. Thus, while it
appears that the power in the high energy component is greater than
that in the low energy component, this need not necessarily be the
case since a low energy peak was not measured. However, this would be
a very unusual GRB spectrum, as peak energies are typically between 100
keV and 1 MeV~\cite{preece}.

The existence of a higher energy component of emission is
predicted by many emission models. For models in which the sub-MeV
emission is due to non-thermal emission from a population of
relativistic electrons, a fraction of these photons can be upscattered
to the GeV-TeV range via inverse Compton
scattering~\cite{meszaros93, chiang99, pilla98}. The emergence of this
component is expected to be prompt and coincident with the synchrotron
radiation seen in the keV-MeV range.  Protons may also be accelerated
to very high energies: GeV-TeV gamma rays can be produced by
synchrotron radiation from these protons, or from decaying pions
produced by high energy protons interacting with
photons~\cite{boettcher98}.

The possible VHE detection of GRB 970417a has significant implications
on the distance scale and energy production in the GRB.  VHE gamma
rays interact with lower-energy photons to produce electron-positron
pairs~\cite{gould66,jelley66,primack99,salamon98}.  Thus gamma rays
from distant sources may suffer significant attenuation on the
intergalactic background radiation fields as they traverse the
Universe en route to Earth. There is an energy dependent horizon
beyond which gamma rays cannot be detected.  The magnitude of this
effect is a function of gamma-ray energy, the density and spectrum of
the background radiation fields, and the distance to the source.  The
observation of photons with energies of at least 650 GeV coupled with
an estimate of the opacity for VHE photons would constrain the
redshift of GRB 970417a.  The results of Primack et al (1999) imply
that the opacity of the Universe to 650 GeV photons is one at a
redshift of $\sim$0.1.  While it is possible that GRB 970417a may lie
beyond this, it would then require an enormous source flux of VHE
photons to produce the excess observed by Milagrito.

If the redshift of GRB970417a is less than ~0.1, then with the
exception of GRB980425 it is much closer than all of the GRBs with
measured redshifts. However only long, bright bursts have been
localized sufficiently well to allow the measurement of afterglows, so
the sample of GRBs with measured redshifts may not be representative
of the entire class of GRBs. While GRB970147a was a long burst, it was
also dim and would not have been localized by past or current
detectors.

There have been several attempts to extend information on the GRB
redshift distribution to bursts without measured redshifts.
Several authors have investigated the gamma-ray properties of bursts
with known redshifts (and thus luminosities) to find observable
gamma-ray parameters that may be indicators of the luminosity. The
spectral evolution of pulse structures appears to be anticorrelated
with peak luminosity (i.e. bursts with a long lag between low-energy
and high-energy detection by BATSE are dimmer)~\cite{norris00} and
bursts with a greater degree of variability appear to be more
distant~\cite{fenimore01}.  Although GRB970417a was too dim at sub-MeV
energies to allow the calculation of the variability parameter to be
used as a luminosity indicator, it was included in a recent study of
the lag-luminosity relation~\cite{norris02}. However, because
GRB970417a was both dim and of relatively short duration, the large
uncertainty in the measured lag precludes obtaining a reliable
estimate of its luminosity~\cite{norris02a}.  Other distance
indicators are obtained from the global properties of GRBs due to
deviations from Euclidean geometry for sources at cosmological
distances.  For example, Schmidt (2001) shows that GRBs with harder
MeV spectra are more distant than those with soft spectra, and in fact the
spectrum of GRB970417a, as measured by BATSE, is soft.

In addition to suffering attenuation via pair production in
intergalactic space, VHE photons may also be absorbed in the source
itself. The very rapid variability (small emission region) and high
luminosity of GRBs implies a very large photon density. If the source
is non-relativistic, the optical depth of high-energy photons is so
large that the photons could not emerge. If the emission region
is moving relativistically, then the pair production optical depth
is decreased because the photon energies and densities in the rest
frame of the emission region are lower than they appear to the
observer~\cite{baring97,lithwick01}. Using the method of (Lithwick and
Sari), we can use the requirement that the burst be optically thin to
650 GeV photons to place a lower limit on the Lorentz factor
($\gamma$) of the expansion. Assuming a variability timescale of 1 second,
a broken power law spectral fit to the BATSE data and a redshift of 0.1
we find a lower limit to the Lorentz factor of 95 for GRB970417a. 
Therefore, unusually large Lorentz factors are not required for
GRB970417a, primarily because the implied luminosity of such a
nearby burst is low resulting in a relatively low photon density.

The future of these studies holds great interest.
Milagro, a more sensitive detector than Milagrito, is
now operating.  SWIFT will launch in 2003 and is expected to detect
and localize several hundred bursts per year.  A large fraction of the
SWIFT detections will have measured distances.  Milagro will observe
these nearby GRBs detected by SWIFT to determine the fraction of GRBs
with TeV emission as well as the flux of that TeV emission.

\acknowledgments

Many people helped bring Milagrito to fruition.  In particular, we
acknowledge the efforts of Scott Delay, Gwelen Paliaga, Neil Thompson
and Michael Schneider.  This work was supported in part by the
National Science Foundation, the U. S.  Department of Energy (Office
of High Energy Physics and Office of Nuclear Physics), Los Alamos
National Laboratory, the University of California, and the Institute
of Geophysics and Planetary Physics.


\begin{thebibliography}{99}

\bibitem[Atkins et al. 2000a]{atkins00a} Atkins, R.~et al.\ 2000a, 
Nuclear Instruments and Methods in Physics Research A, 449, 478 

\bibitem[Atkins et al. 2000b]{atkins00b} Atkins, R.~et al.\ 2000b, 
\apjl, 533, L119 

\bibitem[Atkins et al. 1999]{atkins99} Atkins, R.~et al.\ 1999, 
\apjl, 525, L25 

\bibitem[Band et al. 1993]{band93} Band, D.~et al.\ 1993, 
\apj, 413, 281. 

\bibitem[Baring \& Harding 1997]{baring97} Baring, M.~G.~\& 
Harding, A.~K.\ 1997, \apj, 491, 663 

\bibitem[Boettcher \& Dermer 1998]{boettcher98} Bottcher, M.~\& 
Dermer, C.~D.\ 1998, \apjl, 499, L131 

\bibitem[Briggs et al. 1999]{briggs99} Briggs, M. S., Pendleton, 
G. N., Kippen, R. M. , Brainerd, J. J., Hurley, K. , Connaughton, V.  \& 
Meegan, C. A. 1999, \apjs, 122, 503 

\bibitem[Chiang \& Dermer 1999]{chiang99} Chiang, J.~\& Dermer, 
C.~D.\ 1999, \apj, 512, 699 

\bibitem[Connaughton 1999]
{connaughton99} Connaughton, V. 1999 (private communication) 

\bibitem [Dermer\comma Chiang \& Mitman 1999]
{dermer99} Dermer, C.D., Chiang, J., \& Mitman, K.E. 1999, \apj (submitted)



\bibitem[Fenimore 2001]
{fenimore01} Fenimore, E. at al. 2001, \apj, 552, 57

\bibitem[Fishman 1995]{fishman95} Fishman, G.~J.\ 1995, \pasp, 
107, 1145 

\bibitem[Gould \& Schreder 1966]
{gould66}Gould, R.J. \& Schreder, G. 1966, \prl, 16(6), 252

\bibitem[Helene 1983]{helene83} Helene, O., et al. 1983, Nucl. Instr. Meth. 
Phys. Res. 212, 319


\bibitem[Hurley et al. 1994]{hurley94} Hurley, K., et al. 1994, \nat, 372, 652

\bibitem[Jelley 1966]
{jelley66}Jelley, J.V. 1966, \prl, 16(11), 479

\bibitem[Kaneko 2000]
{kanekoy00} Kaneko, Y. (private communication) 




\bibitem[Lithwick \& Sari 2001]{lithwick01} Lithwick, Y.~\& Sari, 
R.\ 2001, \apj, 555, 540 

\bibitem[Meszaros \& Rees 1993]{meszaros93} Meszaros, P.~\& Rees, 
M.~J.\ 1993, \apjl, 418, L59 


\bibitem[Norris 2000]{norris00} Norris, J. 2000, \apj, 534, 248

\bibitem[Norris 2002]{norris02} Norris, J. 2002, astro-ph 0201503

\bibitem[Norris 2002a]{norris02a} Norris, J. 2002, private communication.
                       



\bibitem[van Paradijs\comma Kouveliotou \& Wijers 2000]{paradijs00} 
van Paradijs, J., Kouveliotou, C., \& Wijers, R.~A.~M.~J.\ 2000, 
\araa, 38, 379

\bibitem[Pilla \& Loeb 1998]{pilla98} Pilla, R. P. \& Loeb, A.  
1998, \apjl, 494, L167 


\bibitem[Preece et al. 2000]{preece} Preece, R.~D., Briggs, 
M.~S., Mallozzi, R.~S., Pendleton, G.~N., Paciesas, W.~S., \& Band, D.~L.\ 
2000, \apjs, 126, 19 

\bibitem [Preece 1999]
{preece99} Preece, R. 1999, (private communication)

\bibitem[Primack\comma Bullock\comma Somerville \& MacMinn 1999]{primack99} 
Primack, J. R., Bullock, J. S., Somerville, R. S. \& MacMinn, D.  1999, 
Astroparticle Physics, 11, 93 

\bibitem[Salamon \& Stecker 1998]{salamon98} Salamon, M. H. \& 
Stecker, F. W. 1998, \apj, 493, 547 

\bibitem[Schmidt 2001]{schmidt01} Schmidt, M.\ 2001, \apj, 552, 
36 

\bibitem[Schneid et al. 1992]{schneid92} Schneid, E.~J.~et al.\ 1992, \aap, 255, L13

\bibitem[Schneid et al. 1995]{schneid95} Schneid, E.~J.~et al.\ 1995, \apj, 453, 95

\bibitem[Sommer et al. 1994]{sommer94} Sommer, M.~et al.\ 1994, \apjl, 422, L63


\end{thebibliography}
\end{document}